\documentclass[aps,prl,reprint,groupedaddress,amsmath,amssymb,aps]{revtex4-2}
\usepackage{graphicx}
\usepackage{dcolumn}
\usepackage{bm}
\usepackage{floatrow}

\begin{document}


\title{Multiobjective optimization in design of broadband extreme ultraviolet multilayers}


\author{Shang-qi Kuang}
\email[]{ksq@cust.edu.cn}
\author{Xiao-wei Song}
\author{Jing-quan Lin}
\email[]{linjingquan@cust.edu.cn}
\affiliation{School of Science, Changchun University of Science and Technology, Changchun 130022, China}%



\date{\today}

\begin{abstract}
In order to design broadband extreme ultraviolet multilayers with many objectives, the multiobjective genetic algorithm and the multiobjective genetics algorithm with reference direction have been improved and combined used. The certain conflicting relations between three primary design objectives have been obtained by analyzing the distribution of nondominated solutions, and the multilayer designs with critical average reflectivities have been found. Basing on the multiobjective genetics algorithm with reference direction, the exact multilayer design has been obtained by guiding the searching in desired region of solution space. Our method can supply the multilayer designs which are the optimal trade-offs among these design objectives, and it has a great potential in designing optical multilayers with more objectives.
\end{abstract}


\maketitle

Owing to the demanding applications in extreme ultraviolet (EUV), accelerator based light sources, soft X-ray microscopy, and astronomy telescopes, the multilayer optics have experienced significant development over the last three decades \cite{Pirozhkov:2015, Huang:2017}. The narrow spectral and angular reflectivity bandpass of EUV mulrilayer that results from the interference nature of EUV reflection from periodic multilayer limit the applications of multilayer optics \cite{Zameshin:2019}. A straightforward way to increase the reflected bandwidth is to use the aperiodic multilayer, and this method have been well developed and intensively investigated in this field \cite{Yakshin:2010, Kuang:2017}. Recently, it is a routine method to develop a broadband EUV multilayer via consideration of the conventional performance of reflected profile, and much effort has been made towards to meet some other advanced requirements in multilayer design, such as the throughput of reflected bandpass, relative dispersion of reflectivity \cite{Kozhevnikov:2015}, effect of random thickness errors \cite{Yakshin:2010}, polarization degree of reflected plateau \cite{Tang:2017} and so on. Unfortunately, in most of these researches, those advanced performances are test using the multilayer structure which is obtained by only optimizing the reflected profile, which is unable to optimize those advance performances. Furthermore, the merit function which describes those advanced performance has been added with traditional merit function, and then the summation has been optimization \cite{Kozhevnikov:2015, Yang:2016}. Though this method is simple to convert the optimized objectives into a single goal, but it fails to analyze the complex relations between different design goals, and one must give a good consideration of weight between the objectives. It is also realized that the aimed reflectivity is a determined constant in previous approaches of multilayer designs, which can not explore the potential performances or risks of multilayer system, and then the throughput and flatness of reflected plateau should be optimized as two different design targets. Furthermore, as the industrial development of EUV optics \cite{Bakshi:2018}, the influence of random thickness errors has to be an additional objective. Therefore, further developments in EUV optics make it necessary to simultaneously optimize many objectives of multilayer design, and how to obtain the best possible trade-offs is becoming a fundamental challenge. 

Fortunately, many multiobjective genetic algorithms (MOGA) have been proposed and performed well on optimizing several objectives simultaneously \cite{Lin:2018}. Due to the conflict among all these objectives, this optimization generates a set of solutions representing the best possible trade-offs, which compose the Pareto-optimal set, and the mapping of Pareto-optimal set in objective space is defined as Pareto-optimal front. Furthermore, in order to find a single preferred solution in the evolutionary multiobjective optimizations efficiently, the MOGA using reference direction has been developed \cite{Liu:2018}. However, the multilayer design requires to optimize many parameters and the merit functions are complex, thus the solution space is quite large, and it is not easy to obtain the Pareto-optimal front. Up to now, MOGA has been used in the designs of broadband EUV multilayers with two optimized targets \cite{Kuang:2019}. In this letter, we improved an MOGA and developed the MOGA with angle-based preference selection mechanism (MOGA-ANGLE), and then we combined these two algorithms and expanded them into the designs of EUV broadband multilayers with 3D objectives.

The multilayer mirror with a wide angular bandpass is an important component for large aperture EUV optical system, and its reflected plateau is very sensitive to layer thickness errors \cite{Yakshin:2010}. Therefore, its multilayer design is very representative, and our approach to the inverse problem of designing EUV multilayer with multiobjectives is based on the minimization of these merit functions 
\begin{equation}
\begin{aligned}
&f_{1}=\left (\int_{\theta_{\text{min}}}^{\theta_{\text{max}}}R(\theta)d\theta\right)^{-1}=\left (\bar{R}(\theta_{\text{max}}-\theta_{\text{min}})\right)^{-1};\\
&f_{2}=\int_{\theta_{\text{min}}}^{\theta_{\text{max}}}\left(  \frac{R(\theta)}{\bar{R}}-1\right)^{2}d\theta;\\
&f_{3}=f_{2}+\frac{1}{2}\sum_{i=1}^m\frac{\partial^{2}f_{2}}{\partial d^{2}_{i}}\delta_{i}^{2},
\label{eq:one}
\end{aligned}
\end{equation} 
where $R$ and $\bar{R}$ are the theoretical and average reflectivities of the designed multilayer, and $\theta_{\text{min}}$ and $\theta_{\text{max}}$ are the minimum and maximum incident angles, respectively. The first merit function $f_{1}$ is the reciprocal of reflected throughput, and the second merit function $f_{2}$ characterizes the deviation of the calculated reflectivity profile from the average reflectivity. The third merit function $f_{3}$ represents sensitivity of reflected profile to random thickness errors, and here $d_{i}$ and $\delta_{i}$ are the thickness and thickness error's standard deviation of the $i$th layer respectively. We need to optimize the thicknesses of $m$ layers contained in the multilayer system, and these layers have the ineluctable thickness errors which originate from imprecision deposited control of quartz crystal monitoring or time monitoring. Therefore, these three performances of multilayer can be optimized by minimization of these functions, and we set them as optimized objectives of MOGA. The conceptual steps and more details of MOGA with two optimized targets can be found in Ref.\cite{Kuang:2017} and the references therein. Because the dimension of objectives is increased from 2D to 3D, several improvements have been made. At first, the increasing of objective dimension induces an enormous enlargement of solution space, and then we increase the population size to 1000, and run the program until 16000 generations. Secondly, we adopt the archive truncation method \cite{Lin:2018} to keep nondominated solutions of new population based on their Euclidean distances to the nearest neighbor to replace the strategy based on their crowding-distance values, which further enhanced the diversity of MOGA. Thirdly, in order to prevent premature convergence of MOGA that most solutions have very low reflected plateaus, we adopt the penalty functions for the individual, if the value of first merit function $f_{1}$ is larger than 0.15, all its other values of merit functions are revalued by $f_{i}+\beta (i=2,3)$, where $\beta=10^{6}$. This strategy can exclude the individual with low reflected plateau ($\bar{R}\leq 42\% $) from the first nondominated front. It is worthwhile to point out that the MOGA is well suited for parallelization at multi-processor high-performance computing systems.

Although, the MOGA mentioned above can be used to find a set of representative efficient solutions of multilayer designs in solving multiobjective optimization, the nondominated solutions obtained are hard to near the Pareto front, and it is difficult to derive exact solutions in local solution space. Therefore, we further introduce the MOGA with reference direction in the multilayer designs. In order to ensure the consistency between MOGA and MOGA with a reference direction, their basic steps and strategies are the same, but the crowding-distance and archive truncation method are instead of angle-based preference selection mechanism (MOGA-ANGLE) \cite{Liu:2018}. We chose the representative efficient solution $\vec{F_{r}}=(f_{1r},f_{2r},f_{3r})$ in the nondominated solutions and provide this reservation point as the searching direction, and then for the individual $j$ in the population, we connect it to the aspiration point so as to form a vector and calculate the angle between its vector $\vec{F_{j}}=(f_{1j},f_{2j},f_{3j})$ and the searching direction as \cite{Liu:2018}
\begin{equation}
\Theta=\arccos\left(\frac{\vec{F_{j}}\cdot\vec{F_{r}}}{\left|\vec{F_{j}}\right|\cdot\left|\vec{F_{r}} \right|}\right)
\label{eq:two}.
\end{equation}   
As a result, this approach not only makes the selection pressure stronger but also integrates the preference information of aspiration point to guide the search toward desirable region in solution space. 

The improved approach is applied to the designs of Mo/Si multilayer mirror, which has a flat reflected profile at wavelength of 13.5 nm in the incidence angular from 0$^{\circ}$ to 16$^{\circ}$ \cite{Yakshin:2010}. In our simulation, the structure of multilayer system can be written as Sub/Si/[MoSi$_{2}/$Mo/MoSi$_{2}$/Si]$_{49}$/SiO$_{2}$, where the imperfections of interface, the interlayers and oxidation of top layer are all considered. For simplicity and without loss generality, we consider $s$ polarization of light, and the SiO$_{2}$ oxide layer has a thickness of 2nm and a surface roughness of 0.6nm r.m.s. Therefore, our multilayer system considered can be an appropriate model for the Mo/Si multilayer deposited by DC magnetron sputtering \cite{Pardidi:2016}. 

\begin{figure}[h]
	\centering\includegraphics[width=7cm]{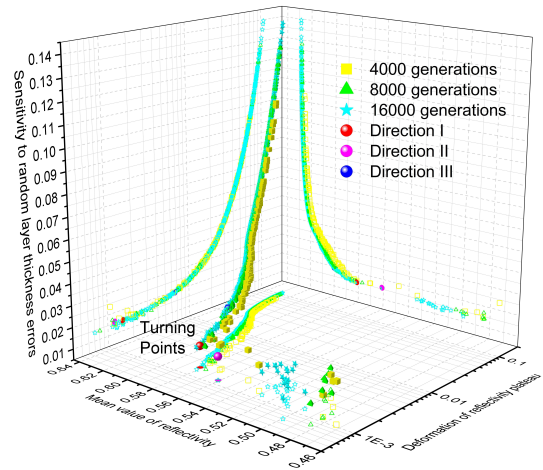}
	\caption{\footnotesize{Obtained nondominated solutions according to different generations of multiobjective genetic algorithm where three merit functions of multilayer designs are considered. Three representative efficient solutions are chosen as the desirable searching directions for the multiobjective genetic algorithm with reference direction.}}\label{fig_1}	
\end{figure}
After the optimizations of MOGA, the nondominated solutions according to different generations are presented in Fig.1, where the value of merit function $f_{1}$ has been converted to mean value of reflectivity. In Fig.1, it is found that these values of three merit functions can be optimized simultaneously, and all individuals populate in the first nondominated front after 14000 generations. It is clear that many solutions with different average reflectivities can be obtained in one run, and this is a great advantage than traditional design method that the aimed reflectivity is set as a constant, and only one multilayer design can be obtained in one simulation. Meanwhile, the relations between different objective are easy to identify and analyze. In Fig.1, with increasing the average reflectivity, the deformation of reflectivity plateau and sensitivity to random thickness errors also increase, which means these objectives conflict. An interesting phenomenon is found that when the solution's average reflectivity is higher than about 58\%, both of deformation of reflectivity plateau and sensitivity to random thickness errors increase dramatically with increasing the average reflectivity; while for the solutions having a lower reflectivities than 58\%, they have comparable deformation of reflectivity plateau and sensitivity to random thickness errors, thus these solutions with critical average reflectivities work as the turning points. This result can be understood by the reflected ability of multilayer system, and this information is very important for multilayer designs to reduce the deposition risks of EUV mirrors. However, because the solution space is very large, it is found that we have not obtained many solutions having an acceptable deviation of the reflectivity cure from a flat plateau. Therefore, we chose three representative solutions in Fig.1 as searching directions to search the desirable solution regions, and here the chosen solutions of Directions I (locate in the region of turning points), II and III have the lower sensitivity to random thickness errors, smaller deviation from a flat plateau and relative higher average reflectivity, respectively.    

\begin{figure}[h]
	\centering\includegraphics[width=7cm]{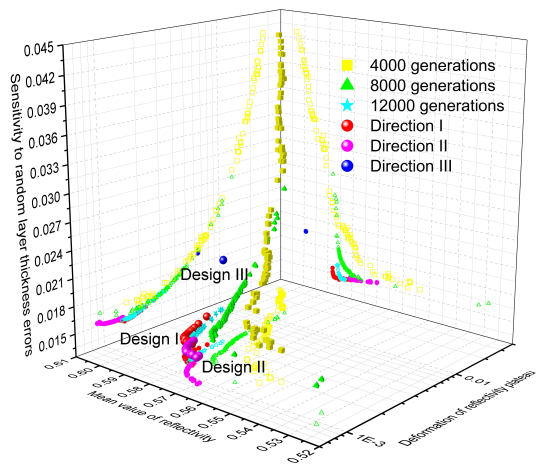}
	\caption{\footnotesize{Obtained nondominated solutions according to different generations of multiobjective genetic algorithm with reference direction, where the Direction I as shown in Fig.1 is used searching direction. Obtained nondominated solutions of multiobjective genetic algorithm with reference direction after 16000 generations, and these three sets of multilayer designs are according to three searching directions as shown in Fig.1, respectively. Here, the solutions with lowest sensitivity to random thickness errors, smallest deviation from a flat plateau and highest average reflectivity are defined as Designs I, II and III, respectively.}}\label{fig_2}	
\end{figure}
With the chosen directions as shown in Fig.1, the obtained nondominated solutions based on MOGA-ANGLE are demonstrated in Fig.2. For MOGA-ANGLE, the population size of 100 is enough for local searching, which speeds up the solving process. In Fig.2, the nondominated solutions according to different generations are presented, and it is found that these three merit functions can also be optimized simultaneously. Furthermore, one can see that all the individuals populate in the first nondominated front very quickly, and all solution concentrate in the region of Direction I, which means the local searching has been realized. Meanwhile, these nondominated solutions after 16000 generations according to Directions I, II and III are demonstrated in Fig.2, respectively. It is realized that further optimization has obtained induced by MOGA-ANGLE, the solutions with more stability of reflectivity plateau to thickness errors and more flat reflected profile are obtained in Directions I and II respectively, and we define them as Designs I and II as shown in Fig.2. After the optimization of 16000 generations in Direction III, all solutions converge to a point, and it is found that all the solutions in Directions are nearly the same, and then we define them as Design III. Therefore, the MOGA and MOGA-ANGLE can complement each other in multilayer design, MOGA can supply the global information of solutions, and MOGA-ANGLE can guide the searching to desired local region to obtain solutions which are closer to Pareto-optimal front.

\begin{figure*}[ht]
	\centering
	\includegraphics[width=13.2cm]{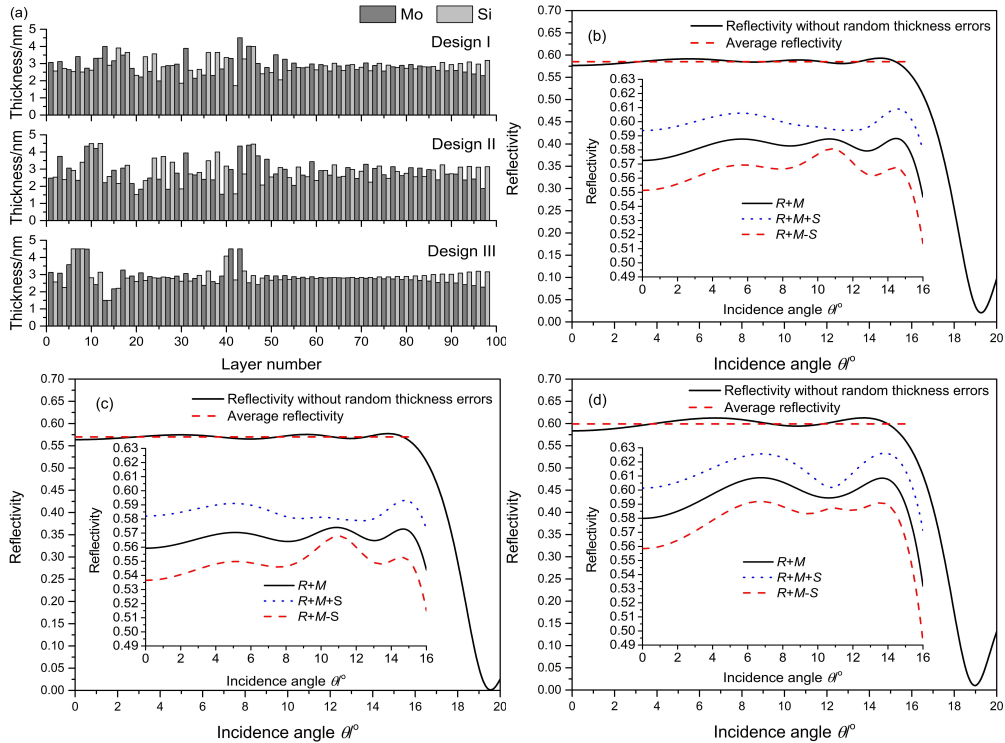}
	\caption{\footnotesize{(a) Designed layer thickness distributions of Mo/Si multilayers according to the multilayer designs as shown in Fig.2, respectively. The naturally formed interlayers are considered, and the thicknesses of the Mo-on-Si and Si-on-Mo interfaces are 1.0nm and 0.5nm, respectively, but these interlayers are not shown; The theoretical reflectivity plateaus and average reflectivities of reflectivity plateaus for Design I (b), Design II (c) and Design III (d), respectively. These insets are the mathematical expectation reflectance $R+M$ and the standard deviation corridors $R+M \pm S$ for the corresponding multilayer designs, respectively. Here the densities of all materials are the same as their bulk densities, the interfacial roughness is 0.3nm r.m.s, and the random thickness errors of Mo or Si layers have the normal distribution and a standard deviation of 0.05nm.}}
	\label{fig_3}	
\end{figure*} 

In Fig.3(a), we demonstrate the layer thickness distributions of Designs I, II and III, respectively, and it is easy to found that these multilayer designs are completely different, so our approach can supply a set of multilayer designs. In order to consider the influences of random thickness errors on reflectivity profile, the mathematical expectation reflectivity profile and standard deviation corridor of multilayer design can be given by \cite{Kuang:2017}
\begin{equation}
\begin{aligned}
&M(\theta)=\frac{1}{2}\sum_{i=1}^{98}\frac{\partial^{2}R(\theta)}{\partial d^{2}_{i}}\delta_{i}^{2};\\
&S^{2}(\theta)=\sum_{i=1}^{98} \left( \frac{\partial R(\theta)}{\partial d_{i}} \right )^{2} \delta_{i}^{2}+\frac{1}{4} \sum_{i,j=1}^{98} \left( \frac{\partial ^{2} R(\theta)}{\partial d_{i} \partial d_{j}} \right )^{2} \delta_{i}^{2} \delta_{j}^{2},
\label{eq:three}
\end{aligned}
\end{equation}    
where $d_{j}$ and $\delta_{j}$ are the thickness and thickness error's standard deviation of the $j$th layer, and $M(\theta)$ and $S(\theta)$ are deviation and variation of mathematical expectation reflectivity. The mathematical expectation reflectivity profiles and standard deviation corridors of Designs I, II and III are given in Figs.3(b), 3(c) and 3(d), respectively. Comparing Fig.3(b) and Fig.3(c), one can found that although Design II has a reflected plateau which is more flat, but its average reflectivity is lower and under the standard deviations of 0.05nm thickness errors, the region of perturbed reflection is larger, which means its deposited risk is higher. Meanwhile, Design III can give the highest average reflectivity, but 
an increase of average reflectivity by 2\% induces a great increase in the deviation from a flat plateau and the region of perturbed reflection. As result, Design I can supply the best comprehensive performances, thus the information of turning points is very useful and important. Furthermore, it is very difficult to obtain multilayer structure of Design I for traditional design methods, where the robust design is seldom considered and solution searching is a kind of blind. Most EUV researches focused on developing the high-cost manufacturing technology, but a design improvement can provide improved performances using the affordable technologies in the laboratory. 

In conclusion, the improved MOGA and MOGA-ANGLE are combined and applied in the designs of broadband EUV multilayers with many conflicting objectives. Basing on the optimization of MOGA, the relations between objectives of multilayer design can be found in the first nondominated front, and some critical solutions have been obtained. Three interested solutions are picked out and set as searching directions for MOGA-ANGLE, and then the searches are guided toward the desirable regions. As a result, several competitive results of broadband multilayer designs have been obtained. It is important that the establish of multilayer design with 3D objectives is a critical milestone of designing multilayers with more objectives, thus this work completely opens the new ways of optical thin film design based on multiobjective optimization algorithms.

\begin{acknowledgments}
This project is supported by the National Natural Science Foundation of China (Nos. 61405189 and 61974142) and Jilin Scientific and Technological Development Plan (Nos. 20200401052GX and 20190201013JC).
\end{acknowledgments}

\end{document}